# Coherent detection of the oscillating acoustoelectric effect in graphene

Yicheng Mou[1], Jiayu Wang[1], Haonan Chen[1], Yingchao Xia[1], Hailong Li[2], Qing Yan[1,2,4], Xue Jiang[3], Yijia Wu[1,4], Wu Shi[1,5], Hua Jiang[1,4], X. C. Xie[1,2,4,6], Cheng Zhang[1,5*]

[1]*State Key Laboratory of Surface Physics and Institute for Nanoelectronic Devices and Quantum Computing, Fudan University, Shanghai 200433, China*

[2]*International Center for Quantum Materials, School of Physics, Peking University, Beijing 100871, China.*

[3]*Department of Biomedical Engineering, School of Information Science and Technology, Fudan University, Shanghai 200433, China*

[4]*Interdisciplinary Center for Theoretical Physics and Information Sciences (ICTPIS), Fudan University, Shanghai 200433, China*

[5]*Zhangjiang Fudan International Innovation Center, Fudan University, Shanghai 201210, China*

[6]*Hefei National Laboratory, Hefei 230088, China*

\* Correspondence and requests for materials should be addressed C. Z. (E-mail: zhangcheng@fudan.edu.cn).

**Abstract**

**In recent years, surface acoustic waves (SAWs) have emerged as a novel technique for generating quasiparticle transport and band modulation in condensed matter systems. SAWs interact with adjacent materials through piezoelectric and strain fields, dragging carriers in the direction of wave propagation. Most studies on the acoustoelectric effect have focused on the collective directional motion of carriers, which generates a steady electric potential difference, while the oscillating component from dynamic spatial charge modulation has remained challenging to probe. In this work, we report the coherent detection of oscillating acoustoelectric effect in graphene. This is achieved through the coherent rectification of spatial-temporal charge oscillation with electromagnetic waves emitted by interdigital transducers. We systematically investigate the frequency and gate dependence of rectified signals and quantitatively probe the carrier redistribution dynamics driven by SAWs. The observation of oscillating acoustoelectric effect provides direct access to the dynamic spatial charge modulation induced by SAWs through transport experiments.**

Surface acoustic waves (SAWs) are highly-directional mechanical vibrations that propagate on the surface of elastic materials. Leveraging their unique ability to interact with materials through piezoelectric and strain fields, SAWs have found diverse applications in the field of condensed matter physics. By generating localized electric fields and mechanical deformations, SAWs can modulate band structures, induce quasiparticle transport, and facilitate the exploration of new physical phenomena [1]. Their ability to precisely control carrier dynamics makes them valuable for investigating transport properties, quantum coherence, and nonlinear effects in various materials. The development of transfer methods further offers the opportunity to provide efficient SAW generation for acoustoelectric (AE) effect research by building heterostructures of two-dimensional materials with high-performance piezoelectric substrates such as $LiNbO_3$ [2]. The versatility of SAWs extends to applications in the development of advanced electronic [3–5], optoelectronic [6–9] and spintronic [10,11] devices, where they play a critical role in enhancing device performance and functionality through dynamic modulation of material properties.

SAWs interact with electrons in semiconductors through piezoelectric and strain fields, giving rise to the AE effect [2,12–20]. For SAWs propagating through a piezoelectric substrate along $x$-axis, it creates an accompanying piezoelectric wave in which the in-plane part can be expressed as $E_p(x,t) = E_{p0}e^{i(qx-\omega t)}$ ($q$ for acoustic wave vector). The piezoelectric wave gives rise to a spatial-temporal dependent current in the semiconductor channel, comprising both $d.c.$ and $a.c.$ components, as depicted in Fig. 1(a). Here we refer them as steady and oscillating AE effect, respectively. The steady AE effect manifests as a steady electrical potential difference generated by the directional motion of carriers along the SAW propagation direction, while the oscillating AE effect arises from the oscillatory motion of carriers around their equilibrium positions. This charge modulation, in turn, generate internal electric fields that partially screen the original piezoelectric fields, resulting in acoustic attenuation and velocity shift [21–28]. While traditional studies have predominantly focused on the steady AE effect, the oscillating AE component, inherent to the wave nature of SAWs, may offer deeper insights into carrier dynamics. Probing the oscillating behavior and redistribution of electrons under SAWs is crucial for both fundamental research and the development of acoustoelectronic devices. However, due to the inherent spatial-temporal oscillations, directly observing this transient charge redistribution in transport experiments presents a significant technological challenge.

In this work, we report the detection of oscillating AE effect by coherent rectification of spatial-temporal charge oscillation using electromagnetic waves (EMWs) emitted by interdigital transducers (IDTs). Assuming carriers respond almost instantaneously to the oscillating piezoelectric field [29], a spatial-temporal oscillation of carrier density arises, described by $n(x,t) = n_0 + n_1 e^{i(qx-\omega t)}$. The carrier density oscillation is accompanied by a periodic modulation of the energy band, as illustrated in Fig. 1(b). Based on the continuity equation ($\nabla \cdot \mathbf{j} = -\partial n/\partial t$), $n_1$ can be expressed as:

$$n_1 = \frac{1}{e} \frac{\sigma_{xx}(\omega)/\sigma_m}{1 + i\sigma_{xx}(\omega)/\sigma_m} \varepsilon_0 (1 + \varepsilon_r) E_{p0}, \quad (1)$$

where $\sigma_{xx}(\omega)$ is $a.c.$ sheet conductivity, $e$ is the elementary charge, and $\sigma_m$ is a constant defined by $\sigma_m = v_{SAW}\varepsilon_0(1+\varepsilon_r)$ with $\varepsilon_r$ the effective relative dielectric coefficient [31]. Then the conductivity also becomes spatial-temporal dependent:

$$\sigma_{xx}(x,t) = \sigma_{xx0} + \frac{\partial \sigma_{xx}}{\partial n}\bigg|_{E_F} n_1 e^{i(qx-\omega t)}. \quad (2)$$

To detect the transient charge redistribution state, we introduce EMWs with oscillating electric field component $E_m = E_{m0}e^{i(kx-\omega t)}$ ($k$ for electromagnetic wave vector) with the same frequency as SAW. Since the wavelength of EMW at SAW frequency (sub-GHz to GHz range) is about meter scale, much larger than the sample size, the spatial variation of $E_m$ can be ignored. Using $j_\alpha(x,t) = \sigma_{xx}(x,t)E_\alpha(x,t)$ ($\alpha = x, y$, corresponding to the two directions of the SAW propagation plane), a rectified $d.c.$ current will be generated by the EMW received in the channel:

$$j_\alpha = \frac{1}{2}\frac{\partial \sigma_{xx}}{\partial n} n_1 e^{iqx} E_{m\alpha 0} e^{i\Delta\phi}. \quad (3)$$

Here $E_{m\alpha 0}$ is the electric field amplitude of EMW component along $\alpha$ direction, and $\Delta\phi$ corresponds to the phase difference between SAWs and EMWs. The real part of $j_\alpha$ is then proportional to $\cos(\Delta\phi)$. This rectified current, excited by the EMWs, constitutes a coherence measurement, effectively enabling coherent detection of the oscillating AE effect. Experimentally, the EMWs can be naturally obtained from the SAW excitation process. When IDTs, which are on-chip metal thin films, are driven with radio-frequency (RF) *a.c.* voltages, they not only generate SAWs on the substrate, but also radiate EMWs into free space, akin to antennas [32–35]. These EMWs share exactly the same frequency as SAWs (see Supplemental Material Chap. II for further discussions). Consequently, the phase difference $\Delta\phi$ can be quantified through the time delay between SAW and EMW propagation. Note that the EMW propagation velocity far exceeds the SAW velocity $v_{SAW}$, $\Delta\phi$ can be simplified as $\Delta\phi = 2\pi f d/v_{SAW}$, with $d$ the distance between IDTs and electrodes on the channel. Thus $\Delta\phi$ can be effectively tuned by adjusting the SAW frequency $f$ within the passband of IDTs.

Experimentally, we employ graphene on piezoelectric Y-cut LiNbO$_3$ (LNO) to test the above proposal. Owing to the low carrier density and high mobility properties, graphene serves as an ideal candidate for studying AE effect. Figure 2(a) shows the schematic of graphene AE device used in this study. We fabricated IDTs on LNO substrates by photolithography and metal sputtering, which were designed to generate SAW propagating along the crystal's *z*-axis with a wavelength of 12 μm. The hBN/graphene/hBN trilayer structures were stacked and transferred on LNO within the aperture of the IDTs by dry-transfer method [36]. Then the graphene samples were etched into Hall-bar geometry and edge-contacted with electrodes. The LNO substrate also served as the back gate dielectric. Sound-absorbing tapes were attached to the edge of LNO to eliminate the reflections of SAW. The *S*-parameters of IDTs were examined using vector network analyzer after calibration. Figure 2(c) shows the result of reflection coefficient $S_{11}$ versus frequency at 300 K and 10 K. The resonant frequency at 300 K and 10 K is 282 MHz and 287 MHz, respectively. The inset shows the temperature dependence of the resonant frequency, resulting from the SAW velocity shift with temperature.

The transport measurements were performed in a commercial variable temperature insert with superconducting magnet (1.6-300 K, ±12 T). We used lock-in amplifiers to measure longitudinal voltages and transverse voltages in both electric and AE transport experiments to avoid the influence of contact resistance. In electric transport, sinusoidal current with an amplitude of 0.1 μA and a frequency of 3-17.777 Hz was applied between the source and drain electrodes. The conductivity and Hall mobility of a typical graphene device is $\sigma_{xx} = 3.551\times10^{-3}$ S and $\mu = 2.155\times10^4$ cm$^2$/(V·s) at 300 K and zero gate voltage, indicating a high sample quality. In AE transport measurements, a signal generator was used to input RF power with a sub-gigahertz frequency to excite IDTs through coaxial cables. The AE signals were measured by either lock-in amplifier using amplitude-modulated RF excitation or *d.c.* voltmeters using continuous-wave RF excitation.

Figure 2(d) shows the measured transverse AE voltage $V_{yx}^{AE}$ versus frequency under continuous-wave excitation at 300 K and 0 T with an input RF power of -10 dBm (0.1 mW). Here *y* means the voltage direction, and *x* refers to the SAW propagating direction. We find that $V_{yx}^{AE}$ oscillates around zero as a function of *f*, forming a wave-packet-shape envelope. The envelope shape is proportional to (1-$S_{11}$) and reaches its maximum at the resonant frequency $f_r$, which further supports the acoustic origin of $V_{yx}^{AE}$. Such behavior is in stark contrast to conventional steady AE effect, whose sign is determined by the carrier type and does not change with *f*. On the contrary, these behaviors nicely match the coherent rectification picture of oscillating AE effect proposed above. The oscillation of $V_{yx}^{AE}$ results from the slight variation of SAW wavelength at different frequency, which modifies the phase difference $\Delta\phi$. Given that $\Delta\phi = 2\pi f d/v_{SAW}$, the oscillation periodicity of $V_{yx}^{AE}$ is directly proportional to the distance $d$ between IDTs and sample electrodes.

To quantitatively analyze $V_{yx}^{AE}$, we carried out simulations of $E_{p0}$ and $E_{m0}$ to estimate $V_{yx}^{AE}$ versus frequency as shown in Fig. 2(d). Since $\sigma_{xx} \gg \sigma_m$ ($\sigma_m \sim 10^{-6}$ S for LNO), the carrier oscillation $n_1$ induced by SAW (Eqn. (1)) can be approximately expressed as

$$n_1 = -i\frac{1}{e}\varepsilon_0(1+\varepsilon_r)E_{p0}, \quad (4)$$

which is now independent of $\sigma_{xx}$. For systems with one dominant carrier, $\partial\sigma_{xx}(\omega)/\partial n$ corresponds to $e\mu$ with $\mu$ the carrier mobility under high-frequency electric fields. Then the rectified current of oscillating AE effect can be reduced to:

$$j_{yx}^{AE} = \frac{1}{2}\mu\varepsilon_0(1+\varepsilon_r)E_{p0}E_{my0}\cos(\frac{d}{v_{SAW}}2\pi f). \quad (5)$$

The piezoelectric field amplitude $E_{p0}$ versus $f$ can be calculated from the SAW intensity through [22,23,37]

$$\frac{1}{4}\varepsilon_0(1+\varepsilon_r)E_{p0}^2 v_{SAW} = \frac{K^2}{4}qI_{SAW}. \quad (6)$$

Experimentally, the SAW intensity can be obtained as 0.25 W/m at the center frequency 282 MHz through $S_{11}$ by $I_{SAW} = (1-S_{11})P_{RF}/W$ with $W$ the aperture of IDT. Here $K^2 = 0.046$ and $\varepsilon_r = \varepsilon_{33} = 28.7$ for Y-Z LNO at 300 K [38,39], and the input $P_{RF} = 0.1$ mW. By using finite element analysis, we calculate the distribution of radiation field based on the geometric structure of IDTs and extract $E_{my0}$ to be 64.9 V/m at 282 MHz. Despite potential deviations in simulations due to the complex environmental geometry in reality, the numerical results generally align well with the experimental data, as shown in Fig. 2(d).

Figure 2(e) shows $V_{yx}^{AE}$ as a function of input RF power at a series of frequencies. We track the power dependence of $V_{yx}^{AE}$ at different maximum and minimum positions, which present consistent linear dependence. According to Eqn. (5), $j_{yx}^{AE} \propto \sqrt{P_{SAW}P_{EMW}}$. Since both $P_{SAW}$ and $P_{EMW}$ are proportional to $P_{RF}$, it gives $j_{yx}^{AE} \propto P_{RF}$ as observed in Fig. 2(e). Figure 2(f) shows the 2D diagram of $V_{yx}^{AE}$ as functions of frequency and temperature. The observed shifts in the maxima and minima of the oscillation nicely matches the trend of the center frequency shown in Fig. 2(c). These results indicate that the coherent detection phenomenon is generally robust and observable across a wide temperature range.

To directly distinguish the oscillating AE effect from conventional steady AE effect, we employ a pulsed-wave technique. We turn the continuous-wave RF excitation to the pulsed-wave excitation by adding a pulse switch between the signal generator and IDT. The pulse switch was driven by a TTL voltage, which precisely controls the pulse width ($\tau$) and duty ratio. If the pulse width $\tau$ is shorter than $d/v_{SAW}$, SAWs and EMWs are separated in the time domain as illustrated in Fig. 3(a). Hence, they cannot interfere with each other, resulting in the absence of coherent rectification condition for detecting oscillating AE effect. In this case, the measured AE voltages are mainly from steady AE effect. By increasing $\tau$, the coherent rectification gradually restores owing to the overlap of SAWs and EMWs in time domain (Fig. 3(b)). Experimentally, we measured both the longitudinal AE voltage ($V_{xx}^{AE}$) and transverse AE voltage ($V_{yx}^{AE}$) under pulsed-wave excitation. The RF power was set to 0 dBm (1 mW). The time delay $\Delta t$ was estimated to be around 232 ns based on the distance between IDTs and graphene in our device. Figures 3(c) and 3(d) show $V_{yx}^{AE}$ and $V_{xx}^{AE}$ versus $f$ under pulsed SAW with different pulse widths but a fixed duty ratio of 1:10, respectively. When the pulse width reduces to 0.1 μs, no oscillation was detected in both $V_{xx}^{AE}$-$f$ and $V_{yx}^{AE}$-$f$ curves. The longitudinal AE voltage $V_{xx}^{AE}$ shows a peak at $f_r$, while the transverse one, $V_{yx}^{AE}$, is nearly zero since no magnetic field is applied. They fit well with the steady AE effect scenario as widely studied in earlier works [2,18,19,40]. By gradually increasing $\tau$ from 0.1 to 1 μs, large oscillations of AE signals appear in the frequency domain, similar to the result in Fig. 2. For $V_{xx}^{AE}$, both the steady AE and the coherent rectified AE components can be observed, since both the directional motion and charge oscillation induced by SAW are present in the longitudinal direction simultaneously. The insert of Fig. 3(d) presents the normalized oscillation amplitude versus pulse width $\tau$. We can see that the oscillation amplitude starts to appear at around 232 ns, rapidly increases with $\tau$, and eventually saturates. The oscillation of AE signals converges to the result of continuous-wave condition at $P_{RF}$ = 0.1 mW (considering the duty ratio of 1:10) when $\tau$ is large enough.

We further explore the rectified voltages of oscillating AE effect in graphene at different gate voltages. Figure 4(a) presents the transfer curve of channel resistance by sweeping back-gate voltage with clear ambipolar behavior. The insert of Fig. 4(a) is the carrier density obtained by Hall effect. Figure 4(b) shows $V_{yx}^{AE}$ versus $V_{BG}$ at different frequencies. We see that the sign of $V_{yx}^{AE}$ changes as the dominant carrier type changes regardless of the frequency. It is because the sign of mobility $\mu = v/E$ is different for electrons and holes, leading to the sign change of $V_{yx}^{AE}$. According to Eqn. (6), the rectified signal of oscillating AE effect is proportional to mobility for one-carrier systems. We calculate $\mu_{AE}$ (which is $\partial\sigma_{xx}(\omega)/\partial n$) as well as mean free path at different gate voltages using simulated electric field amplitude of EMW value ($E_{m0}$ = 64.9 V/m). A comparison of mean free path $\bar{l}_{AE}$ extracted from oscillating AE effect and $\bar{l}_H$ extracted from Hall effect versus electron density is presented in Fig. 4(c), showing similar profile. We further calculate the carrier variation part $n_1$ versus input RF power in Fig. 4(d). The carrier variation gets larger towards the center frequency and systematically increases with $P_{RF}$. It allows a quantitative measure of the carrier redistribution driven by SAWs. These results demonstrate the potential of detecting the oscillating AE effect to reveal critical information about AE transport.

In summary, we have demonstrated a novel method for coherently detecting the oscillating acoustoelectric effect using graphene. By leveraging the interaction between SAWs and EMWs emitted by IDTs, we are able to probe the dynamic spatial charge modulation induced by SAWs. The systematic investigation of the frequency and gate dependence of the rectified signals provides valuable insights into the underlying mechanisms of the oscillating AE effect. This work not only advances our understanding of carrier dynamics in the presence of SAWs but also opens new avenues for exploring dynamic charge modulation in two-dimensional materials and other condensed matter systems. The ability to coherently detect and manipulate these oscillations holds significant potential for the development of advanced electronic and optoelectronic devices.

We gratefully thank Yang Liu, Yuhang Li, Zengwei Zhu, and Haiwen Liu for helpful discussions. This work was supported by the National Key R&D Program of China (Grant No. 2022YFA1405700), the National Natural Science Foundation of China (Grant No. 12174069 and 92365104), and Shuguang Program from the Shanghai Education Development Foundation. Part of the sample fabrication was performed at Fudan Nano-fabrication Laboratory.

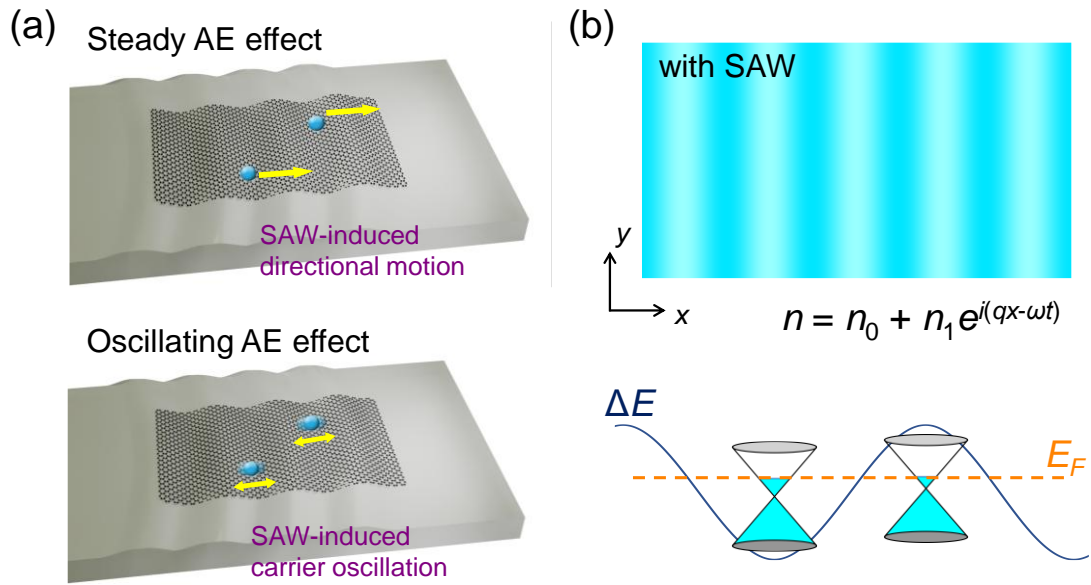

(single column)

FIG. 1. (a) Illustration of the different mechanisms and carrier response of steady and oscillating AE effect. (b) Illustration of dynamic spatial charged carrier modulation (top panel) and energy band modulation (bottom panel) induced by SAWs. The orange dashed line marks the position of Fermi energy, and the blue curve represents the magnitude of position-dependent electric potential.

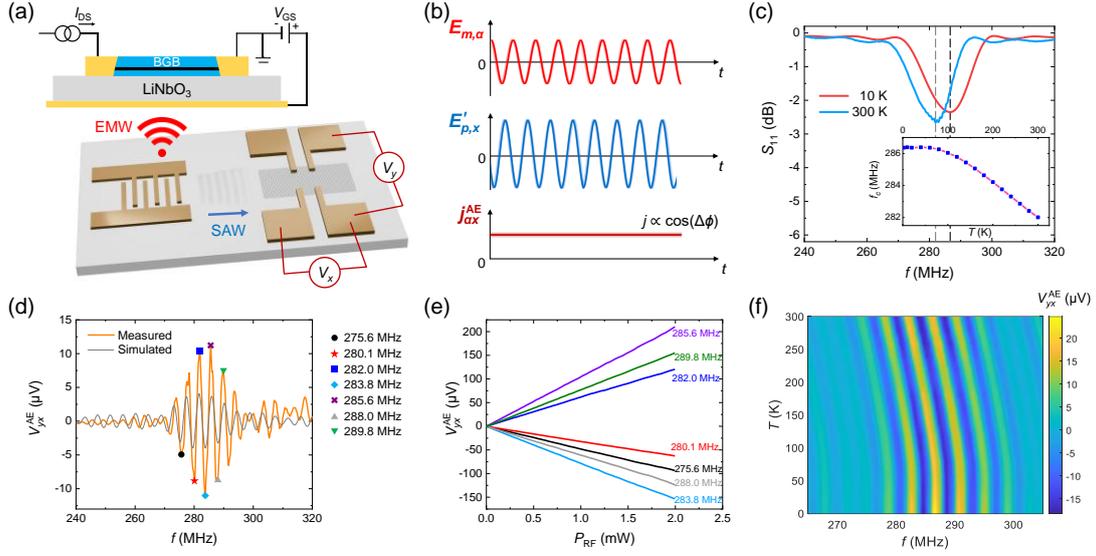

FIG. 2. (a) Schematic illustration of the BN-encapsulated-graphene-stacked SAW devices and the measurement setup (including both electric transport and AE effect measurements). When inputting RF voltages, the IDTs simultaneously generate SAWs and EMWs. (b) Illustration of the time-dependent electric field intensity of EMW ($E_{m,\alpha}$), the screened piezoelectric wave ($E'_{p,x}$) as well as the induced *d.c.* current density $j_{\alpha x}^{AE}$ in the sample. (c) Characterization of IDTs by vector network analyzer: the reflection coefficient $S_{11}$ in logarithmic scale versus the input RF frequency under different temperatures. The results show standard IDT behavior (bandpass filter). The inset shows the resonant frequency versus temperature extracted from $S_{11}$. (d) The measured and simulated transverse AE voltage in frequency domain at 300 K under continuous-wave mode. The RF power is -10 dBm (0.1mW). The frequencies corresponding to maximum/minimum values are marked in the figure. (e) Power-dependent transverse AE voltage at different frequencies where $V_{yx}^{AE}$ reaches its maximum/minimum. (f) 2D diagram of $V_{yx}^{AE}$ as a function of frequency and temperature. The RF power is -10 dBm under continuous-wave mode.

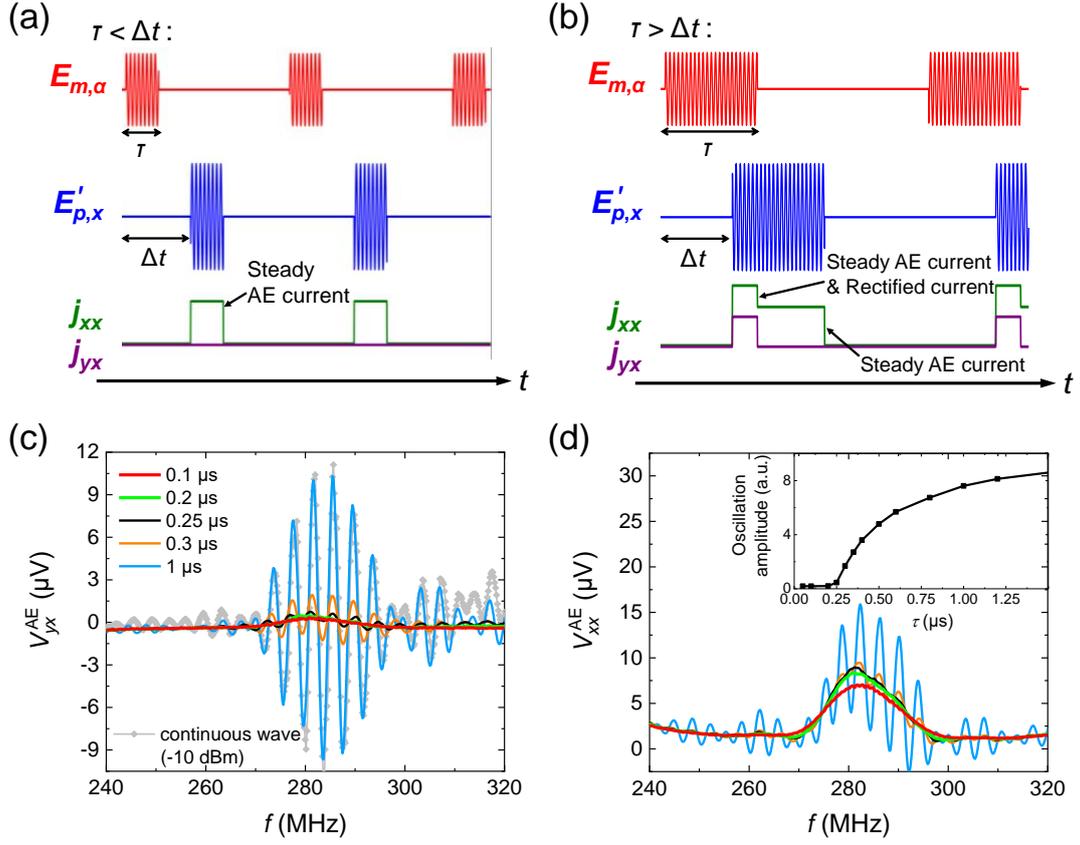

(single column)

FIG. 3. (a)-(b) Illustration of the induced transverse ($j_{yx}$) and longitudinal ($j_{xx}$) currents in the pulse-SAW experiments under different pulse widths. If the EMW and SAW pulses do not overlap in time, the measured currents revert to the ordinary steady AE effect, with $j_{yx}$ vanishing. If the EMW and SAW pulses partial overlap, additional oscillating AE currents arise when the two waves coexist. (c) Transverse AE voltage under pulsed SAW with a fixed duty ratio (1:10) but different pulse widths. The RF power is 0 dBm. Oscillating signals only arises when $\tau > \Delta t$. As $\tau$ increases, $V_{yx}^{AE}$ gradually enhances and transitions to the result of continuous-wave condition. (d) Longitudinal AE voltage under pulsed SAW with different pulse widths with $P_{RF}$ = 0 dBm (1 mW). $V_{xx}^{AE}$ does not vanish when $\tau < \Delta t$ due to the presence of steady AE effect. The insert figure shows the normalized oscillating part amplitude versus pulse width.

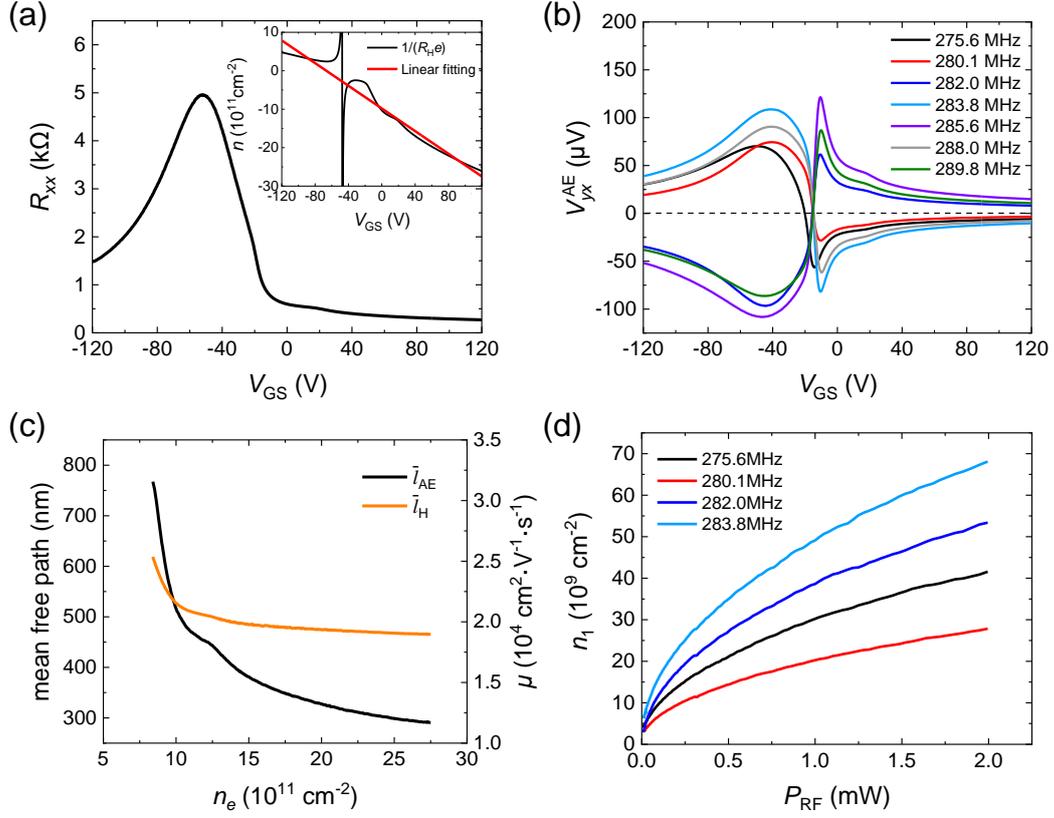

(single column)

FIG. 4. (a) Transfer curve over gate voltage at 300K reveals standard bipolar behavior of graphene. The insert shows the gate-dependent carrier density obtained by measuring Hall resistivity at $B = \pm 1$ T. (b) Transverse AE voltage versus gate voltage at several frequencies. (c) Mobility and mean free path comparison of $\bar{l}_H$ and $\bar{l}_{AE}$ versus electron density. The $\bar{l}_H$ is calculated from Hall resistance and $\bar{l}_{AE}$ is extracted by the measured rectified AE voltage at center frequency. (d) The carrier variation part $n_1$ versus input RF power at a series of frequencies.